\documentclass[a4paper]{article}
\usepackage{INTERSPEECH2022,cite,comment,amssymb}
\let\OLDthebibliography\thebibliography
\renewcommand\thebibliography[1]{
  \OLDthebibliography{#1}
  \setlength{\parskip}{1pt}
  \setlength{\itemsep}{1pt plus 0.3ex}
}
\title{Improving Voice Trigger Detection with Metric Learning}
\name{Prateeth Nayak$^1$, Takuya Higuchi$^1$, Anmol Gupta$^2{}^{\ast}$, Shivesh Ranjan$^1$, Stephen Shum$^1$, Siddharth Sigtia$^1$, Erik Marchi$^1$, Varun Lakshminarasimhan$^1$, Minsik Cho$^1$, Saurabh Adya$^1$, Chandra Dhir$^3{}^{\ast}$, Ahmed Tewfik$^1$ 
}
\address{
  $^1$Apple\\
  $^2$Department of Computer Science, The University of Hong Kong\\
  $^3$JPMorgan Chase $\&$ Co.}
\email{prateethvnayak@apple.com}

\begin{document}

\maketitle
\begin{abstract}
%Voice trigger detection is an important task which enable to activate voice assistant by just speaking a keyword phrase.  For a personal device such as a smart phone, it is also important to ensure that the keyword phrase is spoken by a user of a device, which is called personalized voice trigger detection. Personalized voice trigger detection is typically performed by a cascade system of a speaker independent voice trigger detector and a speaker verification system.
Voice trigger detection is an important task, which enables activating a voice assistant when a target user speaks a keyword phrase. A detector is typically trained on speech data independent of speaker information and used for the voice trigger detection task.
However, such a speaker independent voice trigger detector typically suffers from performance degradation on speech from underrepresented groups, such as accented speakers. In this work, we propose a novel voice trigger detector that can use a small number of utterances from a target speaker to improve detection accuracy. Our proposed model employs an encoder-decoder architecture. While the encoder performs speaker independent voice trigger detection, similar to the conventional detector, the decoder is trained with metric learning and predicts a personalized embedding for each utterance. A personalized voice trigger score is then obtained as a similarity score between the embeddings of enrollment utterances and a test utterance. The personalized embedding allows adapting to target speaker's speech when computing the voice trigger score, hence improving voice trigger detection accuracy. Experimental results show that the proposed approach achieves a 38\% relative reduction in a false rejection rate (FRR) compared to a baseline speaker independent voice trigger model.
\end{abstract}

\section{Introduction}
\renewcommand{\thefootnote}{\fnsymbol{footnote}}
\footnotetext[1]{Work performed at Apple}
\renewcommand*{\thefootnote}{\arabic{footnote}}
% Voice trigger detection is an important task, which enables activating a voice assistant by just speaking a keyword phrase. For personal devices, such as  smart phones, it is also important to ensure that the keyword phrase is spoken by the owner of the device by running a speaker verification system.
Voice trigger detection for personal devices, such as smart phones, is an important task which enables activating a voice assistant by speech containing a keyword phrase. It is also important to ensure that the keyword phrase is spoken by the owner of the device by running a speaker verification system.

A typical approach is to cascade speaker independent voice trigger detection and speaker verification  \cite{jia20212020,hou2021npu,rikhye2021personalized,8683669}. A universal voice trigger detector is trained on speech signals from various speakers to perform speaker independent voice trigger detection, then speaker verification is performed by a speaker recognition model exploiting enrollment utterances spoken by the target user. Various approaches have been proposed for speaker independent voice trigger detection including ASR-based approaches \cite{319341,miller2007rapid,zhuang2016unrestricted,rosenberg2017end,he2017streaming,adya2020hybrid}, as well as discriminative approaches with convolutional neural networks (CNNs)\cite{sainath2015convolutional,tang2018deep,choi2019temporal,majumdar2020matchboxnet}, recurrent neural networks (RNNs) \cite{fernandez2007application,sun2016max,arik2017convolutional,khursheed2021tinycrnn} and attention-based networks \cite{adya2020hybrid,berg2021keyword}.  However, such speaker independent voice trigger detectors typically suffer from performance degradation on speech from underrepresented groups such as accented speakers \cite{shor2019personalizing,viglino2019end}. This is true even when a small amount of adaptation data is available, since adapting a large  speaker independent voice trigger detector is a challenge with only limited data.
%Although a small amount of adaptation data, i.e., the enrollment utterances for the speaker verification system, is available, adaptation of the speaker independent voice trigger detector with such a limited amount of data is challenging.

%In this work, we propose a novel approach that can use enrollment utterances to perform fast adaptation of the voice trigger detector and reduce the number(s) of false rejections and/or false positive activations.
In this work, we propose a novel approach for fast adaptation of the voice trigger detector to reduce the number(s) of false rejections and/or false positive activations. Our proposed model consists of an encoder that performs speaker independent voice trigger detection and a decoder that performs speaker-adapted voice trigger detection. The decoder summarizes acoustic information in an utterance and produces a fixed dimensional embedding. The model is trained using metric learning, where we maximize distance between embeddings of a keyword phrase and non-keyword phrases. We also minimize distance between embeddings of a keyword phrase spoken by the same speaker, and maximize the distance of those spoken by different speakers. The metric learning encourages the model to learn not only differences between the keyword and non-keywords, but also those between keyword phrases spoken by different speakers, thus enabling speaker adaptation. At test time, a speaker-adapted voice trigger score can be obtained as the distance between speaker-specific embeddings extracted from previously seen utterances and embeddings from a test utterance.
%At test time, a speaker-adapted voice trigger score can be obtained as a distance of embeddings computed from enrollment utterances and a test utterance. 
%The encoder and the decoder are jointly trained with a multi-task learning (MTL) framework combining a voice trigger dataset and a speaker recognition dataset.
%A final voice trigger score is obtained by combining a speaker independent voice trigger score from the encoder and the personalized score from the decoder, which boosts the voice trigger detection performance.

Experimental results show that the proposed approach achieves a 38\% relative reduction in a false rejection rate (FRR) compared to a baseline speaker independent voice trigger model for a voice trigger detection task.
%Moreover, for a personalized voice trigger detection task, our proposed approach is on par with a baseline cascaded approach with ??\% relative model size reduction, and achieves a ??\% relative reduction in FRRs when combined with a speaker verification model.

\section{Related work}
\label{sec:related}
Query-by-example is a popular approach for keyword spotting that can also exploit enrollment utterances  \cite{5372889,zhang2009unsupervised,anguera2013memory,kim2019query,7178970,9003781,huang2021query}. In this approach, an acoustic model converts an audio input into a useful representation, e.g., phonetic representation, and then a similarity between the representations of the enrollment and a test utterance is computed using a technique such as dynamic time warping \cite{5372889,zhang2009unsupervised,anguera2013memory} or finite-state transducers \cite{kim2019query}. Phrase-level embedding computed by neural networks is also used as the representation in recent work \cite{7178970,9003781,huang2021query}. Our proposed approach efficiently integrates the essence of the query-by-example approach with the speaker independent voice trigger detector using an encoder-decoder architecture. 
Moreover, speaker-aware training is performed in our approach using metric learning to explicitly differentiate between keyword phrases from speakers and non-keyword speech from same or non-target speakers.
%Our ablation study shows that speaker-aware training improves the voice trigger detection performance (see section \ref{sec:exp}).
%In addition, the speaker independent acoustic models are used in the query-by-example approach and no speaker-aware training is performed to explicitly model differences between speakers and reject keyword phrases from non-target speakers. Our ablation study shows that speaker-aware training improves the voice trigger detection performance (see section \ref{sec:exp}).

Regarding joint modeling for voice trigger detection and speaker verification, Sigtia et al.\cite{9054760} used multi-task learning (MTL) and trained a single model with two branches for voice trigger detection and speaker verification, respectively. Our proposed approach is an extension of \cite{9054760} by adding extra training objectives to reject non-keyword phrases spoken by a target speaker. Note that a simple speaker verification system cannot suppress non-keyword speech from the target speaker, and thus cannot be used for improving voice trigger detection accuracy.

Acoustic model adaptation can also be performed by feeding a speaker embedding into the acoustic model along with audio features \cite{6639211,6707705,6853591}. The speaker embedding can be computed by running a speaker identification model on the enrollment utterances. In contrast, we compare embeddings of known utterances and test utterances for voice trigger detection as we aim to detect whether the two utterances contain the same content, i.e., the keyword phrase, spoken by the same speaker. 

\begin{figure}
    \includegraphics[width=0.45\textwidth]{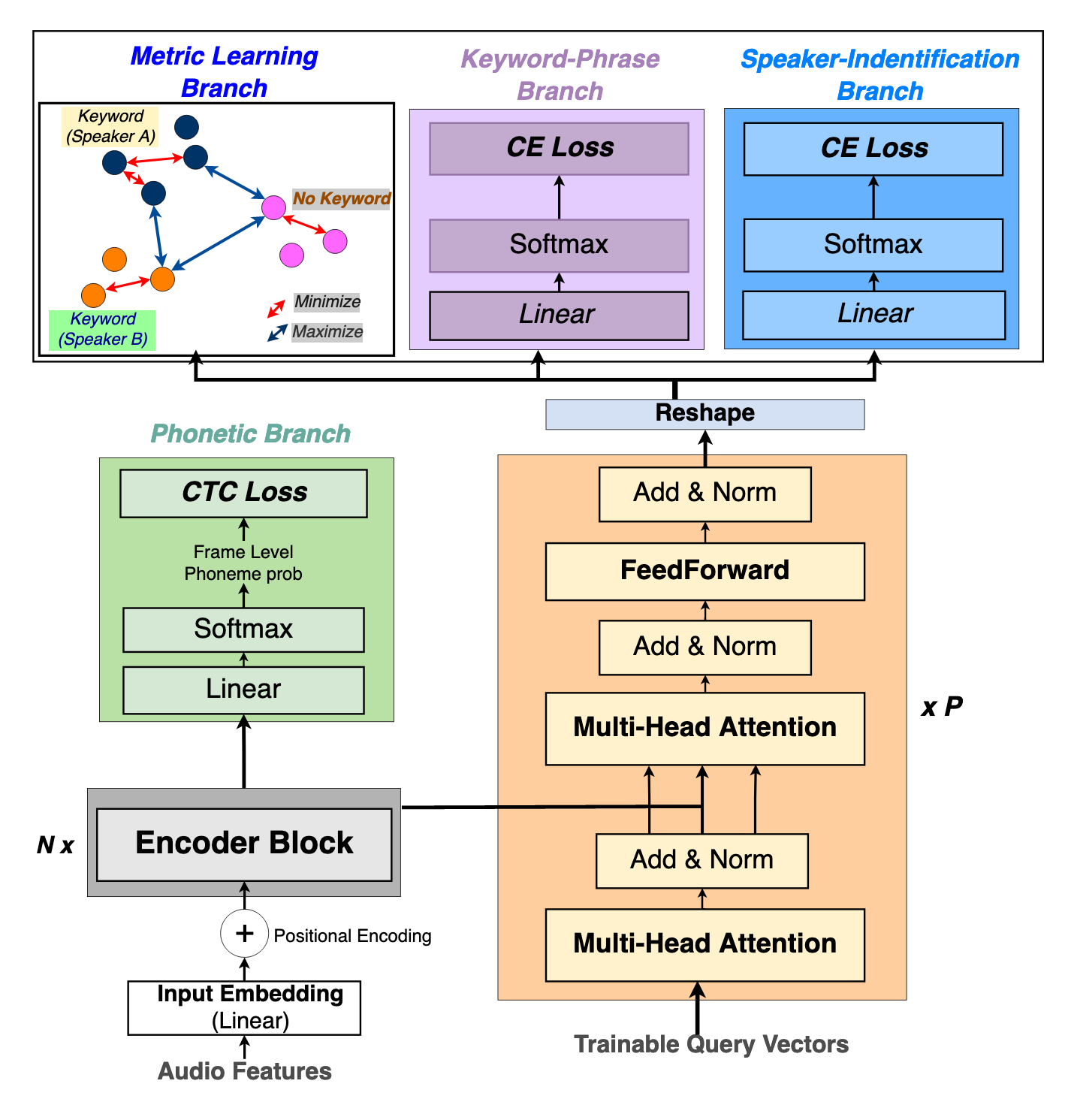}
    \caption{\textit{Proposed MTL Framework: Phonetic Encoder (grey) and Cross-Attention Decoder (orange) blocks. Green block is the Phonetic Branch with CTC Loss ($\mathcal{L}^\textit{(phone)}$). Blue block is the Speaker-Identification branch with CE Loss ($\mathcal{L}^\textit{(spkr)}$). Purple block denotes the Keyword-Phrase Branch with CE Loss ($\mathcal{L}^\textit{(phrase)}$) and, Metric Learning branch ($\mathcal{L}^\textit{(metric)}$).}}
    \label{fig:arch}
    %\vspace{-3mm}
\end{figure}
\section{Proposed approach}
%In this work, we propose a novel approach that can exploit the enrollment utterance $\mathbf{x}'$ to predict the presence of the keyword more precisely.  
We propose a novel MTL approach where an encoder performs a speaker independent phoneme prediction, and a decoder performs speaker-adapted voice trigger detection. 
%The encoder and the decoder are trained jointly in a multi-task learning framework. 
See Figure \ref{fig:arch} for an overview of our proposed approach.

%Since the enrollment is an audio example of the keyword phrase spoken by the target speaker, our proposed approach simply compares the enrollment and test utterances in the embedding domain rather than performing model adaptation using the enrollment utterance or speaker embedding. Additionally, our approach is able to employ metric learning in the embedding domain to differentiate the audio samples originating from a target speaker to that of an imposter or, between two samples originating from the same target speaker in terms of presence or absence of the keyword phrase. We use the cosine distance as the metric to perform this separation between the embedding obtained at the decoder. This adds as an additional branch in the MTL framework along with the keyword-phrase detection task and speaker identification task at the decoder as shown in figure \ref{fig:arch}.

\subsection{Model architecture}
We borrow the model architecture from \cite{9687967} and adapt it for speaker-adapted voice trigger detection. 
The model is based on an encoder-decoder\cite{vaswani2017attention} Transformer architecture. 
%The model employs an encoder-decoder architecture with a phonetic encoder and a cross-attention decoder.
Our encoder consists of $N$ stacked Transformer encoder blocks with self-attention. The self-attention encoder performs phoneme predictions which transforms the input feature sequence, i.e., denoted by $\textit{X}$,  into hidden representations as 
\begin{equation}
\mathcal{I}_\textit{1},\mathcal{I}_\textit{2},...,\mathcal{I}_\textit{N} = \textit{Encoder}(\textit{X}),
\end{equation}
where $\mathcal{I}_\textit{n}$ denotes a hidden representation after the $n$-th encoder block.
A linear layer is applied to the last encoder output $\mathcal{I}_\textit{N}$ to get logits for phoneme classes which are used to compute a phonetic loss. 
%of the batch sampled from the ASR dataset.

Our cross-attention decoder comprises of Transformer decoder blocks with attention layers. The decoder takes the encoder embedding output after the $n$-th encoder block \begin{math}\mathcal{I}_\textit{n}\end{math} as well as a set of trainable query vectors as inputs. Following \cite{9054760}, we use an intermediate representation ($n < N$) since the speaker information can be diminished at the top encoder layer. Let \begin{math}\mathcal{Q}=\{q_m\|\textit{m}=1,...,M\}\end{math} denote a set of the trainable vectors, where \begin{math}q_m \in \mathbb{R}^{d \times 1}\end{math}. By feeding the encoder output and the query vectors, a set of decoder embedding vectors is obtained as
\begin{equation}
e_\textit{1},e_\textit{2},...,e_\textit{M} = \textit{Decoder}(\mathcal{I}_\textit{n},\mathcal{Q}),
\end{equation}
where \begin{math}e_m \in \mathbb{R}^{d \times 1}\end{math} denotes an output of $P$ stacked Transformer decoder blocks. The set of the decoder outputs is then reshaped to form an utterance-wise embedding vector of size $dM \times 1$. Unlike \cite{9687967} that uses the decoder embedding only for a phrase-level cross entropy loss, we use the embedding for three different losses for speaker-adapted voice trigger detection. We first branch out at this stage into two task level linear layers -- one linear layer is applied on the embedding to predict a scalar logit for the keyword phrase; another linear layer is applied to obtain logits for speaker verification. Finally, we also use the decoder embedding to perform metric learning within a mini-batch.

\subsection{Multi-task learning}
In contrast to the previously-proposed MTL framework for keyword spotting  \cite{panchapagesan2016multi,9053577,9054760,9687967}, we introduce the metric-learning loss, to obtain a speaker-adapted voice trigger detection score by comparing the decoder embeddings. In our proposed MTL framework, the model is trained using the phonetic loss at the encoder output and at the decoder output we have three branches -- keyword-phrase loss, speaker-identification loss and the metric-learning loss. 
%In every mini-batch of \textit{N} samples, we sample \textit{M} utterances from ASR dataset and the \textit{(N - M)} samples from speaker ID dataset. 
The objective function for the training can be formulated as 
\begin{equation} \label{mtl}
\mathcal{L} = \mathcal{L}^\textit{(phone)} + \alpha \mathcal{L}^\textit{(phrase)} + \beta \mathcal{L}^\textit{(spkr)}
+ \gamma \mathcal{L}^\textit{(metric)},
\end{equation}
where \begin{math}\mathcal{L}^\textit{(phone)}\end{math}, \begin{math}\mathcal{L}^\textit{(spkr)}\end{math}, \begin{math}\mathcal{L}^\textit{(phrase)}\end{math} and \begin{math}\mathcal{L}^\textit{(metric)}\end{math} denote the phonetic loss, the speaker-identification loss, the keyword-phrase loss and the metric learning loss, respectively. 
%where \begin{math}\mathcal{L}^\textit{(phone)}\end{math}is the phonetic loss applied on \textit{M} utterances, and \begin{math}\mathcal{L}^\textit{(spkr)}\end{math} specifies the speaker-identification loss applied on \textit{(N - M)} utterances.
%\begin{math}\mathcal{L}^\textit{(phrase)}\end{math} is keyword-phrase loss applied on all the samples in the mini-batch since each sample from ASR and speaker-enrollment dataset also contain a keyword phrase label.  \begin{math}\mathcal{L}^\textit{(metric)}\end{math} is the metric learning loss applied on the decoder embeddings by creating sample pairs using the speaker labels and phrase labels within a batch of utterances. 
\begin{math}\alpha,\beta,\gamma\end{math} are the scaling factors for balancing the losses.

We use a phoneme-level connectionist temporal classification (CTC) loss for the phonetic loss \begin{math}\mathcal{L}^\textit{(phone)}\end{math} to compute a speaker independent voice trigger detection score from the encoder output. The keyword phrase loss \begin{math}\mathcal{L}^\textit{(phrase)}\end{math} is a cross-entropy (CE) loss on the scalar logits obtained from the decoder branch with the utterance-wise phrase labels. Similarly, a speaker CE loss \begin{math}\mathcal{L}^\textit{(spkr)}\end{math} is computed using the other decoder branch which constitutes the speaker-identification loss. The speaker-ID CE loss acts as a regularizers, which help generalize the model (see our ablation study in section \ref{results}).

The metric loss \begin{math}\mathcal{L}^\textit{(metric)}\end{math} is a cosine similarity metric with scale and offset parameters that is applied directly on the decoder embedding output for positive pairs, defined as utterances from same speaker containing the keyword phrase; and the negative pairs constitute utterances from different speakers, or utterances from same speaker with opposite phrase labels (see Fig.\ref{fig:arch}). We first convert the cosine similarity into a probability as
\begin{align}
    P_{ij} = (a \cos{\theta_{ij}} + b + 1) / 2 \label{eq:sim},
\end{align}
where $\cos{\theta_{ij}}$ is a cosine distance between the decoder embeddings of the $i$-th and $j$-th utterances. $a$ and $b$ denote trainable scale and offset parameters, respectively. The metric loss \begin{math}\mathcal{L}^\textit{(metric)}\end{math} can be computed as
\begin{align} \label{metricloss}
    \mathcal{L}^\textit{(metric)} = - \frac{1}{N_{\mathcal{P}}}\sum_{(x_i,x_j) \in \mathcal{P}} log P_{ij} - 
    \frac{1}{N_{\mathcal{N}}} \sum_{(x_k,x_l) \in \mathcal{N}} log(1 - P_{kl}),
\end{align}
where $\mathcal{P}$ and $\mathcal{N}$ denote sets of the positive and negative pairs within a mini-batch, and $N_{\mathcal{P}}$ and $N_{\mathcal{N}}$ denote the numbers of positive and negative pairs. 
%$(x_i, x_j)$ are sampled from a set of the positive pairs $\mathcal{P}$ and $(x_k, x_l)$ are sampled from that of the negative pairs $\mathcal{N}$. $\mathcal{P}$ and $\mathcal{N}$ are defined within a mini-batch.
%, and $\cos{\theta_ij}$ and $cos{\theta_kl}$ are then the distance metric, then the contrastive/metric loss can be formulated as Eq.\ref{metricloss}. 
%We also balance the positive and negative pairs when applying this contrastive loss per mini-batch.
We balance the numbers of positive and negative pairs when computing the loss by randomly sub-sampling the negative pairs. The metric-learning loss computes a speaker-adapted voice trigger score in a consistent way during training and inference.

\subsection{Data Sampling} \label{sampling}
We use two sources of data per mini-batch for training the MTL tasks. The first source is set of anonymized utterances that have either the phoneme labels or keyword phrase labels (voice-trigger data), which is mainly used for the phonetic loss and the keyword phrase loss. Non-keyword utterances from the voice-trigger data are also used for the metric learning loss as a negative class. The dataset can be obtained by combining an ASR dataset with the phoneme labels and a keyword spotting dataset with the keyword phrase labels \cite{9053577,9687967}. The other dataset includes utterances with speaker labels (speaker-ID data), where each utterance contains a keyword phrase followed by a non-keyword sentence. The speaker-ID data are used for all of the losses, except the phonetic loss since there is no transcription for this dataset.

We employ a batch sampling strategy that picks samples from both of these sets for every mini-batch of training. For example, for a batch size of 128, we pick 112 utterances from the speaker-ID data which includes 4 utterances from 28 unique speakers, and the rest comes from the voice-trigger data. Also, we randomly drop the keyword phrase segment for the utterances sampled from the speaker data to create negative pairs (keyword vs non-keyword) for the same speaker, which helps metric learning.
% We do explore the variation in the number of unique speakers and number of utterances per speaker parameters in our ablation experiments and the results are shared in the section\ref{}

\subsection{Inference} \label{inference}
% At inference, we first obtain an anchor embedding as an average of the decoder embeddings from existing utterances for a speaker.
During inference, an anchor embedding is obtained first as an average of the decoder embeddings from existing utterances of a speaker that contain a keyword phrase. Next, we compute the decoder embedding on the test utterance, and then compute the similarity score between the anchor embedding and the test embedding using Eq. (\ref{eq:sim}). The similarity score corresponds to the speaker-adapted voice trigger score ($S_{metric}$). Optionally, we combine the speaker-adapted score and a speaker independent voice trigger score $S_{ctc}$ obtained from the encoder output. First the speaker-adapted score is calibrated as $S_{metric}=(P_{i,anchor}-C)/D$ where $C$ and $D$ are the global mean and standard deviation of the scores computed on a validation set. Then we use a simple weighted average to combine these two voice trigger scores:
\begin{align} \label{eq:inference}
    S_{final} = (1-\mu) * S_{ctc} + \mu * S_{metric},
\end{align}
where $\mu$ is a weight factor.

\section{Experimental evaluation}
\label{sec:exp}
\subsection{Data}

% The training data are 2700 hours of anonymized utterances that are randomly sampled from intentional voice assistant invocation recordings and manually transcribed for phonetic labels (54-dimensional). These audio data are augmented with room-impulse responses (RIRs) and echo residuals to obtain a total of approximately 9 million utterances, similar to \cite{9054760,higuchi2021multi}. We add roughly 65\textit{k} utterances that result in false triggers and 300\textit{k} utterances that result in true triggers that are randomly sampled from anonymous speakers for the keyword phrase detection task. The training data for the speaker identification task comprises 15 million utterances from intentional voice assistant invocations. The set contains 131\textit{k} different anonymized speakers with minimum of 100 samples, and median of 115 random samples per speaker. These contain only speaker labels, and no phonetic information. However, each training utterance contains the keyword phrase and the meta information of keyword phrase segment. The training data are formed by concatenating these datasets and we use the batch sampling strategy mentioned in Section\ref{sampling} to ensure each mini-batch contains samples for all tasks.

The training data are thousand hours of randomly sampled anonymized utterances from recordings and manually transcribed for phonetic labels (54-dimensional). These audio data are augmented with room-impulse responses (RIRs) and echo residuals to obtain a total of approximately 9 million utterances, similar to \cite{9054760,9687967}. We add roughly 65\textit{k} false triggers and 300\textit{k} true triggers that are short-lived anonymized utterances randomly sampled from speakers for the keyword phrase detection task. The training data for the speaker identification task comprises 15 million utterances. The set contains 131\textit{k} different anonymized speakers with minimum of 100 samples, and median of 115 random samples per speaker. These contain only speaker labels, and no phonetic information. However, each utterance contains the keyword phrase and the start-stop information of keyword phrase segment. The training data are formed by concatenating these datasets and we use the batch sampling strategy mentioned in Section \ref{sampling} to ensure each mini-batch contains samples for all tasks.

For evaluation, we use a synthetic dataset, where 7535 positive samples are internally collected under controlled conditions from 72 different speakers, evenly divided between genders. Each utterance contains the keyword phrase followed by a voice command spoken to a smartphone. The acoustic conditions include quiet, external noise from TV or kitchen appliances, and music playback.
%at medium and loud volume. 
%This includes a total of 2 hours with a minimum of 2 utterances and median of 9 utterances per speaker. 
To measure false accept (FA) per hour, we include negative data of 2\textit{k} hours of audio recordings that do not contain keyword phrase by playing podcasts, audiobooks, TV, etc. 
%These negative samples are also recorded through intentional invocation of the same device assistant similar to \cite{higuchi2021multi}.
We randomly sample five utterances per speaker for computing the anchor embedding, and we evaluate using the remains utterances. To estimate the variability, we repeat this five times for each speaker, changing the utterances that are used to compute the anchor embedding.  We report the mean performance over the five runs.
%We randomly sample 5 utterances per speaker to simulate the enrollment utterances and evaluate using the rest. To reduce random variability, we repeat the same procedure 5 times for each speaker by changing the random set of the enrollment utterances and report the mean performance.

Similar to \cite{9687967}, we use a two stage approach to reduce the overall compute cost and accommodate the Transformer-based architecture on device for voice trigger detection. We first run light-weight $5 \times 32$ fully-connected neural networks on continuous audio and obtain audio segments of keyword candidates using hidden Markov model (HMM) alignments. Then only the detected audio segment is fed into the baseline/proposed model and a voice trigger score is recomputed. See \cite{9687967} for more details.

\subsection{Model training}
%We compare our proposed MTL approach with the baseline architecture in \cite{higuchi2021multi}.
We use a speaker independent voice trigger detector proposed in \cite{9687967} as a baseline. The baseline system has an encoder-decoder architecture that is trained with the speaker independent phonetic and keyword phrase losses on the voice trigger data. The input features are 40-dimensional log mel-filter bank features $\pm$ 3 context frames, and sub-sampled once per three frames which reduces computational complexity. We also normalize the features using the global mean and variance. A phonetic encoder has 6 layers of Transformer encoder blocks, where each block of multi-head attention has a hidden dimension of 256 and 4 heads. The feed forward network has 1024 hidden units. The final encoder output is projected into 54-dimensional logits using a linear layer. This encoder is trained with CTC loss using the phonetic labels. 
A decoder consists of one Transformer decoder block with the same hidden dimensions as the encoder.
%The cross attention decoder stack consists original Transformer decoder style configuration with the main addition being the cross-attention block. 
The query vector has dimension $d$ of 256 and length $M$ is fixed to 4. The final decoder output embedding is reshaped into 1024 $(256 \times 4)$ dimensional. The baseline approach has the phrase-level CE loss on decoder output for the keyword phrase detection. We also investigate metric-based inference described in section \ref{inference} even though the baseline model is not trained with the metric-learning loss. 

For our proposed approach, we add another linear layer with a dropout of 0.6 on top of the decoder for the speaker-ID loss with the 131\textit{k} speakers.
%we include the speaker information via additional branch at the decoder output. The final decoder output is fed into another linear layer that projects into logits equal to the total number of unique speakers present in the enrollment dataset for speaker-identification task. In addition, we use vectorized operations on the decoder embedding to perform the cosine similarity for the metric loss.
%Since the speaker-ID dataset does not have phonetic labels,
In addition, we initialize our proposed model with the weights of the baseline model and fix the encoder weights to take advantage of the phonetic performance. We only fine-tune the decoder weights in a transfer-learning fashion with the keyword-phrase CE loss, speaker-identification CE loss and the metric learning loss. Also, we consider the penultimate encoder layer embedding for the decoder input ($n=5$).
%We also explored the architecture by feeding in the output of other intermediate encoder attention layers (i.e Encoder Layers 1-5) as input to the cross-attention decoder and the results are discussed in section\ref{results}.
The scaling factors in Eq.(\ref{mtl}) $\alpha$, $\beta$, $\gamma$ are empirically set to be 1, 1, 0.1, respectively. The optimizer used is \textit{Adam}, where initial learning rate is linearly increased until 0.001 until epoch 2, and then linearly decayed to 0.0007 for the next 25 epochs. We then exponentially decay the learning rate with minimum learning rate of 1e-7 until the last epoch set at 40. We use 64 GPUs for training and the batch size is 128 at each GPU.

%\subsection{Baselines}

%\subsection{t-SNE}

\subsection{Results} \label{results}
\begin{figure}[t]
    \includegraphics[width=0.475\textwidth]{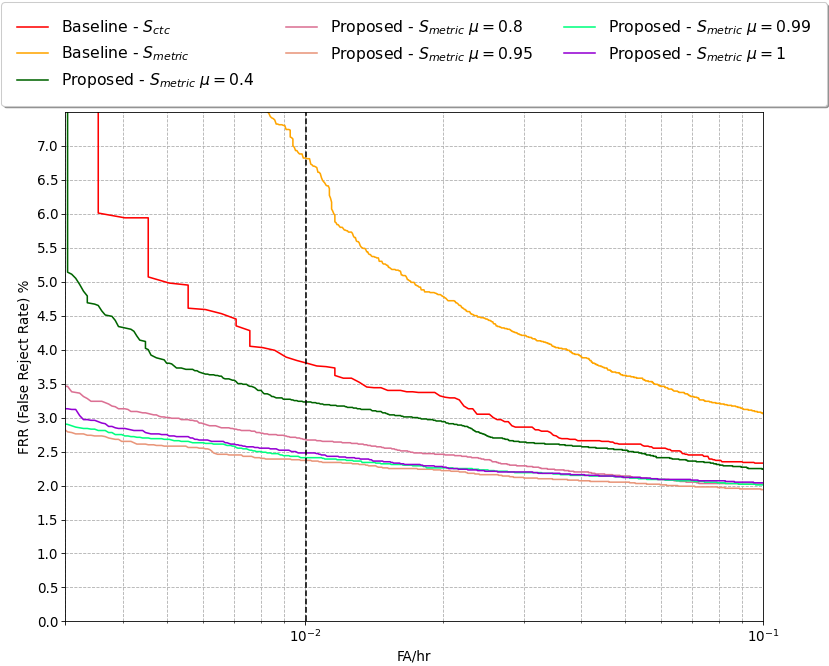}
    %\vspace{-3mm}
    \caption{\textit{DET Curve for evaluation set. The vertical dotted line indicates an operating point.}}
    \label{fig:det}
    \vspace{-3mm}
\end{figure}

Figure \ref{fig:det} shows the detection error trade-off (DET) curves for the baseline and the proposed approach. The horizontal axis represents FA/hr and the vertical axis represents false reject rates (FRRs). Table \ref{tab:FRRs} shows the FRRs at our operating point of $0.01$ FA/hr. The baseline FRR is at 3.8\% when using the phonetic branch for inference. The phrase branch of the baseline shows regressions compared to the phonetic branch even when we apply metric-based inference on the decoder embedding. In addition, fine-tuning the decoder on the speaker-ID data with the  speaker independent phrase loss does not improve the performance.
%We can also see that the baseline performs poorly on the personalized score at 6.82\% FRR since it does not hold any information about the speaker.
In the case of the proposed model,
%the results from both the phonetic branch $S_{ctc}$ and the personalized score $S_{metric}$ are reported, and
we can see that the new MTL improves FRRs. This improvement signifies that the speaker information helps to adapt to the keyword phrase detection using the speaker-adapted score. Additionally, this speaker-adapted score shows the effectiveness of the embedding space being structured by the metric learning. By combining the speaker-adapted score $S_{metric}$ and the speaker independent score $S_{ctc}$ from the phonetic branch,
%As we increase the weight of the personalized score, 
we see further improvement in the FRRs. The proposed approach reduced the FRRs by 38\% relative from the baseline model trained in a speaker independent fashion. 
%In terms of runtime cost and training time, the proposed MTL approach does not add any additional overhead.

\begin{table}[t]
  \caption{False reject rates [$\%$] at an operating point of 0.01 FA/hr.}
  %\vspace{-3mm}
  \label{tab:FRRs}
  \centering
  \scalebox{0.75}{
\begin{tabular}{cccc}
  \toprule
 & Branch & FRRs\\
  \bottomrule
 Baseline \cite{9687967} &\begin{tabular}[c]{@{}c@{}}$S_{ctc}$\\
 $S_{phrase}$ \\ $S_{metric}$ ($\mu$=1)\end{tabular}& \begin{tabular}[c]{@{}c@{}}3.80 \\7.73 \\6.82\end{tabular}\\ 
 + fine-tuning w/ spk-ID data &\begin{tabular}[c]{@{}c@{}} $S_{metric}$ ($\mu$=1)\end{tabular}& \begin{tabular}[c]{@{}c@{}}8.89 \end{tabular}\\ \midrule
 Proposed &\begin{tabular}[c]{@{}c@{}} $S_{ctc}$\\ $S_{metric}$ ($\mu$=1)\end{tabular}& \begin{tabular}[c]{@{}c@{}}3.80 \\\textbf{2.48}\end{tabular}\\ \midrule
 Proposed &\begin{tabular}[c]{@{}c@{}} $S_{ctc}$ and $S_{metric}$ ($\mu$=0.4)\\ $S_{ctc}$ and $S_{metric}$ ($\mu$=0.8)\\ $S_{ctc}$ and $S_{metric}$ ($\mu$=0.95)\\ $S_{ctc}$ and $S_{metric}$ ($\mu$=0.99)\end{tabular}& \begin{tabular}[c]{@{}c@{}}3.23 \\2.67 \\\textbf{2.37} \\2.41\end{tabular}\\ 
    \bottomrule
\end{tabular}
}
\vspace{-3mm}
\end{table}

Table \ref{tab:ablation} shows ablation study for the proposed approach. When we train our proposed model from scratch with encoder layer $n=6$ as decoder input, we see a slight improvement with the metric branch over the baseline. We also see that absence of phonetic loss fails to generalize the model, similar to results reported in \cite{9053577}. Initializing with the baseline model helps retain the phonetic performance $S_{ctc}$, however, the FRR with $S_{metric}$ degrades. This could be because the encoder performs speaker independent phoneme prediction, where speaker information can be diminished at the last layer. Utilising the intermediate encoder layer ($n=5$), we observe improvements.  The rest of Table \ref{tab:ablation} highlights the importance of the three losses on the decoder. We also observe that any fine-tuning of the encoder with the CTC loss introduces performance degradation. 

\begin{table}[ht]
  \caption{Ablation study.}
  %\vspace{-3mm}
  \label{tab:ablation}
  \centering
  \scalebox{0.75}{
\begin{tabular}{cccccccc}
  \toprule
  Init. &
  \begin{math}\mathcal{L}^\textit{(phone)}\end{math} & \begin{math}\mathcal{L}^\textit{(spkr)}\end{math} & \begin{math}\mathcal{L}^\textit{(phrase)}\end{math} & \begin{math}\mathcal{L}^\textit{(metric)}\end{math} & $n$ & Branch & FRRs\\
  \bottomrule
   Random & \checkmark & \checkmark & \checkmark & \checkmark  & 6 & \begin{tabular}[c]{@{}c@{}}$S_{ctc}$\\ $S_{metric}$\end{tabular} &  \begin{tabular}[c]{@{}c@{}}9.00\\3.52 \end{tabular} \\ \midrule
   Random &  & \checkmark & \checkmark & \checkmark  & 6 & $S_{metric}$ &  82.84 \\ \midrule
   %Scratch & \checkmark & \checkmark & \checkmark & \checkmark  & 5 & \begin{tabular}[c]{@{}c@{}}$S_{ctc}$\\ $S_{metric}$\end{tabular} &  \begin{tabular}[c]{@{}c@{}}6.42\\4.86 \end{tabular} \\ \midrule
 Pretrained & Fixed & \checkmark & \checkmark & \checkmark  & 6 & \begin{tabular}[c]{@{}c@{}}$S_{ctc}$\\ $S_{metric}$\end{tabular} &  \begin{tabular}[c]{@{}c@{}}3.80\\28.43 \end{tabular} \\ \midrule
  Pretrained & Fixed & \checkmark & \checkmark & \checkmark  & 5 &  $S_{metric}$ &  \textbf{2.48} \\ \midrule
 Pretrained & Fixed &  & \checkmark & \checkmark & 5 & $S_{metric}$ & 4.53 \\ \midrule
 Pretrained & Fixed & \checkmark & \checkmark & & 5 & $S_{metric}$  & 12.25\\ \midrule
 Pretrained & Fixed &  & \checkmark & & 5 & $S_{metric}$  & 7.50 \\ \midrule
 Pretrained & Fine-tuned & \checkmark & \checkmark & \checkmark  & 5 & \begin{tabular}[c]{@{}c@{}} $S_{ctc}$\\ $S_{metric}$\end{tabular} & \begin{tabular}[c]{@{}c@{}} 7.21\\ 2.66 \end{tabular}\\ 
    \bottomrule
\end{tabular}
}
\vspace{-3mm}
\end{table}

\section{Conclusions}
We propose a novel approach for improving voice trigger detection by adapting to speaker information using metric learning. Our model employs an encoder-decoder architecture, where the encoder performs phoneme prediction for a speaker independent voice trigger detection while the decoder predicts an utterance-wise embedding for speaker-adapted voice trigger detection. The speaker-adapted voice trigger score is obtained by computing a similarity between an anchor embedding for each speaker and the decoder embedding for a test utterance. Experimental results show that our proposed approach outperforms the baseline speaker independent voice trigger detector by $38\%$ in terms of FRRs.

\bibliographystyle{IEEEtran}
\clearpage
\bibliography{mybib}

% \begin{thebibliography}{9}
% \bibitem[1]{Davis80-COP}
%   S.\ B.\ Davis and P.\ Mermelstein,
%   ``Comparison of parametric representation for monosyllabic word recognition in continuously spoken sentences,''
%   \textit{IEEE Transactions on Acoustics, Speech and Signal Processing}, vol.~28, no.~4, pp.~357--366, 1980.
% \bibitem[2]{Rabiner89-ATO}
%   L.\ R.\ Rabiner,
%   ``A tutorial on hidden Markov models and selected applications in speech recognition,''
%   \textit{Proceedings of the IEEE}, vol.~77, no.~2, pp.~257-286, 1989.
% \bibitem[3]{Hastie09-TEO}
%   T.\ Hastie, R.\ Tibshirani, and J.\ Friedman,
%   \textit{The Elements of Statistical Learning -- Data Mining, Inference, and Prediction}.
%   New York: Springer, 2009.
% \bibitem[4]{YourName17-XXX}
%   F.\ Lastname1, F.\ Lastname2, and F.\ Lastname3,
%   ``Title of your INTERSPEECH 2022 publication,''
%   in \textit{Interspeech 2022 -- 23\textsuperscript{rd} Annual Conference of the International Speech Communication Association, September 18-22, Incheon, Korea, Proceedings, Proceedings}, 2022, pp.~100--104.
% \end{thebibliography}

\end{document}